\documentclass{moriondNOTITLES}

\def\be{\begin{equation}}
\def\ee{\end{equation}}
\def\bea{\begin{eqnarray}}
\def\eea{\end{eqnarray}}

\def\fig#1{{Fig.~\ref{#1}}}

\def\vev#1{\left\langle #1\right\rangle}
\def\abs#1{| #1|}

\def\Im{\mbox{Im}\,}

\renewcommand{\bar}{\overline}

\newcommand{\beq}{\begin{equation}}
\newcommand{\eeq}{\end{equation}}

\renewcommand{\(}{\left(}
\renewcommand{\)}{\right)}

\usepackage{amssymb}
\newcommand{\U}{{\rm U}}
\newcommand{\SU}{{\rm SU}}
\newcommand{\SO}{{\rm SO}}

\newcommand{\1}{{\textbf{1}}}
\newcommand{\3}{{\textbf{3}}}

\newcommand{\6}{\textbf{6}}

\usepackage[table]{xcolor}
\definecolor{CGray}{gray}{0.93}

\newcommand{\Photo}{}

\begin{document}
\vspace*{4cm}
\title{Imperfect Axions}

\author{Luca Di Luzio}

\address{Istituto Nazionale di Fisica Nucleare (INFN), Sezione di Padova, \\
Via F.~Marzolo 8, 35131 Padova, Italy}

\maketitle\abstracts{
In this contribution, I'll discuss two classes of effects--additional sources of CP violation and PQ breaking--that offer a slightly different take on axion physics. Both are tied to the idea that the axion solution to the strong CP problem might not be exact, hence the title ``Imperfect Axions''.}

\section{Introduction}

By simultaneously addressing the  strong CP problem  and the dark matter puzzle, the QCD axion 
stands out as a well-motivated 
paradigm, driving the quest for light and weakly-coupled new physics. 
The axion hypothesis, at the interface between particle physics, astrophysics and cosmology, is currently within the reach of
experimental verification. Second generation experiments have started to scratch into the interesting region of parameter space,
while next generation axion experiments are currently being planned. It is conceivable that the axion dark matter mass window could be
thoroughly explored within the next two decades. 

However, nearly half a century after the conception of the Peccei-Quinn (PQ) mechanism \cite{Peccei:1977hh,Peccei:1977ur}, which gives rise to the axion at low energies \cite{Weinberg:1977ma,Wilczek:1977pj}, a compelling theory in which the PQ symmetry emerges naturally is still lacking (PQ origin problem). Moreover, the axion solution is notoriously delicate: ultraviolet (UV) contributions to the axion potential must be extremely suppressed to preserve the solution to the strong CP problem (PQ quality problem).

This contribution explores two classes of effects--additional sources of CP violation and explicit PQ symmetry breaking--that render the axion solution to the strong CP problem ``imperfect'', yet they may imprint interesting phenomenological consequences. I will begin by illustrating how the standard axion mechanism is modified in the presence of such CP-violating (CPV) or PQ-breaking effects, and highlight the resulting phenomenology, in particular axion-mediated forces. I will then turn to the PQ quality problem and introduce a new class of models that address it through the interplay between grand unified theories (GUTs) and flavor symmetries.

\section{Axion dynamics in the presence of CP violation $\&$ PQ breaking}

From an effective field theory (EFT) perspective, the PQ mechanism introduces a new spin-0 field, $a(x)$, endowed 
with a pseudo-shift symmetry, 
$a \to a + \alpha f_a$, 
explicitly broken by an operator that has precisely the same form of the QCD $\theta$-term, 
$\frac{a}{f_a} \frac{\alpha_s}{8\pi}  G \tilde G$.
This symmetry allows $\theta$  to be rotated away by setting $\alpha = -\theta$, thus turning the strong CP problem into a dynamical question about the axion vacuum expectation value (VEV), 
\beq 
\theta_{\rm eff} \equiv \frac{\vev{a}}{f_a} \, , 
\eeq
and the Vafa-Witten theorem \cite{Vafa:1984xg} ensures that the axion VEV relaxes to zero, regardless of the specific form of the axion potential. 
Note, however, that the Vafa-Witten argument  relies on a path-integral inequality and assumes a positive-definite path integral measure, 
which holds for a vector-like theory like QCD but not for weak interactions.

Interestingly, even within the Standard Model (SM), the axion does not relax exactly to zero. Georgi and Randall \cite{Georgi:1986kr} estimated the residual value of $\theta_{\rm eff}$ using naive dimensional analysis. Since $\theta_{\rm eff}$  is a CPV flavor singlet, it shares the quantum numbers of the reduced Jarlskog invariant, $j_{\rm CKM} = \Im V_{ud}V^*_{cd}V_{cs}V^*_{us} \simeq 3 \times 10^{-5}$. To construct such an invariant, one requires the exchange of two $W$ bosons (yielding a  
$G^2_F$ dependence), while long-distance QCD effects account for an extra $\Lambda^4_{\rm QCD}$ factor needed to match dimensions, yielding \footnote{Incidentally, the residual axion dark matter background from misalignment, $\theta^{\rm DM}_{\rm eff} \sim \sqrt{\frac{2 \rho_{\rm DM}}{m_a^2 f_a^2}}$, 
turns out to be of the same size, but for completely unrelated reasons.} 
\beq
\label{eq:thetaeffmin2}
\theta^{\rm SM}_{\rm eff} \sim G_F^2 \Lambda^4_{\rm QCD} j_{\rm CKM}  \sim 10^{-18}  \, .  
\eeq
The resulting estimate, $\theta^{\rm SM}_{\rm eff} \sim 10^{-18}$, 
still far from the sensitivity set by the neutron electric dipole moment (EDM), 
$\abs{\theta_{\rm eff}} \lesssim 10^{-10}$,
is experimentally irrelevant but conceptually important, as it shows that new CPV phases beyond CKM could impact axion dynamics.

Another source of axion imperfections can arise from explicit PQ symmetry breaking. Since $\U(1)_{\rm PQ}$ is a global symmetry, it is expected to be violated by higher-dimensional operators suppressed by some UV scale $\Lambda_{\rm UV}$, 
$\phi^d / \Lambda^{d-4}_{\rm UV} + {\rm h.c.}$, 
where $\phi \sim f_a e^{i a / f_a}$ is the complex field hosting the axion. 
These operators generate additional contributions to the axion potential, generically misaligned with the QCD one. Requiring that this contribution remains subdominant--specifically, less than $10^{-10}$ of the QCD potential--yields
\begin{equation}
\left( \frac{f_a}{\Lambda_{\rm UV}}  \right)^{d-4} f_a^4 \lesssim 10^{-10} \Lambda_{\rm QCD}^4 \, , 
\end{equation}
not to spoil the axion solution to the strong CP problem. 
For example, assuming $\Lambda_{\rm UV} \sim M_{\rm Pl}$  and $f_a \sim 10^9$ GeV, one needs the operator dimension $d > 9$. Thus, the original question of why $\theta$ is small becomes a question of why PQ-breaking operators are so suppressed: the so-called PQ quality problem.

New sources of CP violation beyond CKM can also shift the axion VEV. For instance, consider a short-distance four-quark operator 
$\mathcal{O}_{\rm CPV} = \frac{1}{\Lambda^2_{\rm CPV}} (\bar q q) (\bar q i \gamma_5 q)$, 
suppressed by a UV scale $\Lambda_{\rm CPV}$. In the absence of such operator, the axion potential is dominated by the mass term, set by the QCD topological susceptibility $K = \langle G \tilde G, G \tilde G \rangle \sim \Lambda_{\rm QCD}^4$. When CPV operators are present, the potential acquires a tadpole term, proportional to the correlator $K' = \langle G \tilde G,  \mathcal{O}_{\rm CPV} \rangle \sim \Lambda_{\rm QCD}^6 / \Lambda^2_{\rm CPV}$. Minimizing the potential leads to 
\beq 
\theta_{\rm eff} \equiv - \frac{K'}{K} \sim \frac{\Lambda_{\rm QCD}^2}{\Lambda^2_{\rm CPV}} \simeq 10^{-10} \left( \frac{20 \, {\rm TeV}}{\Lambda_{\rm CPV}} \right)^2 \, , 
\eeq
which saturates the neutron EDM bound for $\Lambda_{\rm CPV} \sim 20$ TeV with $\mathcal{O}(1)$ CPV phases.

In summary, high-scale CP violation or PQ-breaking physics--possibly tied to baryogenesis--can be captured below $\Lambda_{\rm CPV}$ via effective operators, whose contributions to $\theta_{\rm eff}$ can be systematically computed. In addition, such CPV sources can generate a scalar axion-nucleon coupling, leading to axion-mediated forces--an effect we will discuss next.

\section{CP-violating axion couplings}

While standard axion couplings are CP-odd, new sources of CP violation or PQ breaking induce a CP-even scalar coupling to nucleons, $g^S_{aN} \propto \theta_{\rm eff}$.  
Including both terms, the axion interaction Lagrangian  can be written as
\beq 
\label{eq:Laint}
\mathcal{L}^{\rm int}_a \supset 
- g^P_{af} \frac{\partial_\mu a}{2 m_f} \bar f \gamma^\mu \gamma_5 f 
+ g^S_{aN} a \bar N N\, , 
\eeq
where  
$g^P_{af} = C_f m_f / f_a$ (for $f = N, e$), 
with $C_f  \sim \mathcal{O}(1)$ in benchmark axion models \cite{DiLuzio:2020wdo},  
and \cite{Moody:1984ba,Bertolini:2020hjc}
\beq 
\label{eq:fromthetaefftogaN}
g^S_{aN} \simeq \frac{\theta_{\rm eff}}{f_a} \frac{m_u m_d}{m_u + m_d} 
\frac{\langle N | \bar u u + \bar d d | N \rangle}{2} 
\simeq 1.3 \cdot 10^{-12} ~ \theta_{\rm eff} \( \frac{10^{10} \ {\rm GeV}}{f_a} \) \, . 
\eeq 
The scalar axion coupling to nucleons was first discussed by Moody and Wilczek \cite{Moody:1984ba}, and its calculation 
in terms of CPV sources was further developed in subsequent 
works, e.g.~\cite{Pospelov:1997uv,Bertolini:2020hjc,Okawa:2021fto,Dekens:2022gha}. In Ref.~\cite{Bertolini:2020hjc}, 
the calculation of the scalar coupling to nucleons was addressed
within a chiral Lagrangian framework. A key point is that $g^S_{aN}$  receives contributions not only from the axion VEV but also from meson tadpoles. This partially decorrelates the scalar coupling from the neutron EDM, which is important when interpreting experimental constraints. 
A comprehensive analysis, in the context of the SMEFT, can be found Ref.~\cite{Dekens:2022gha}.

\begin{figure}[t]
\centering
\includegraphics[width=14cm]{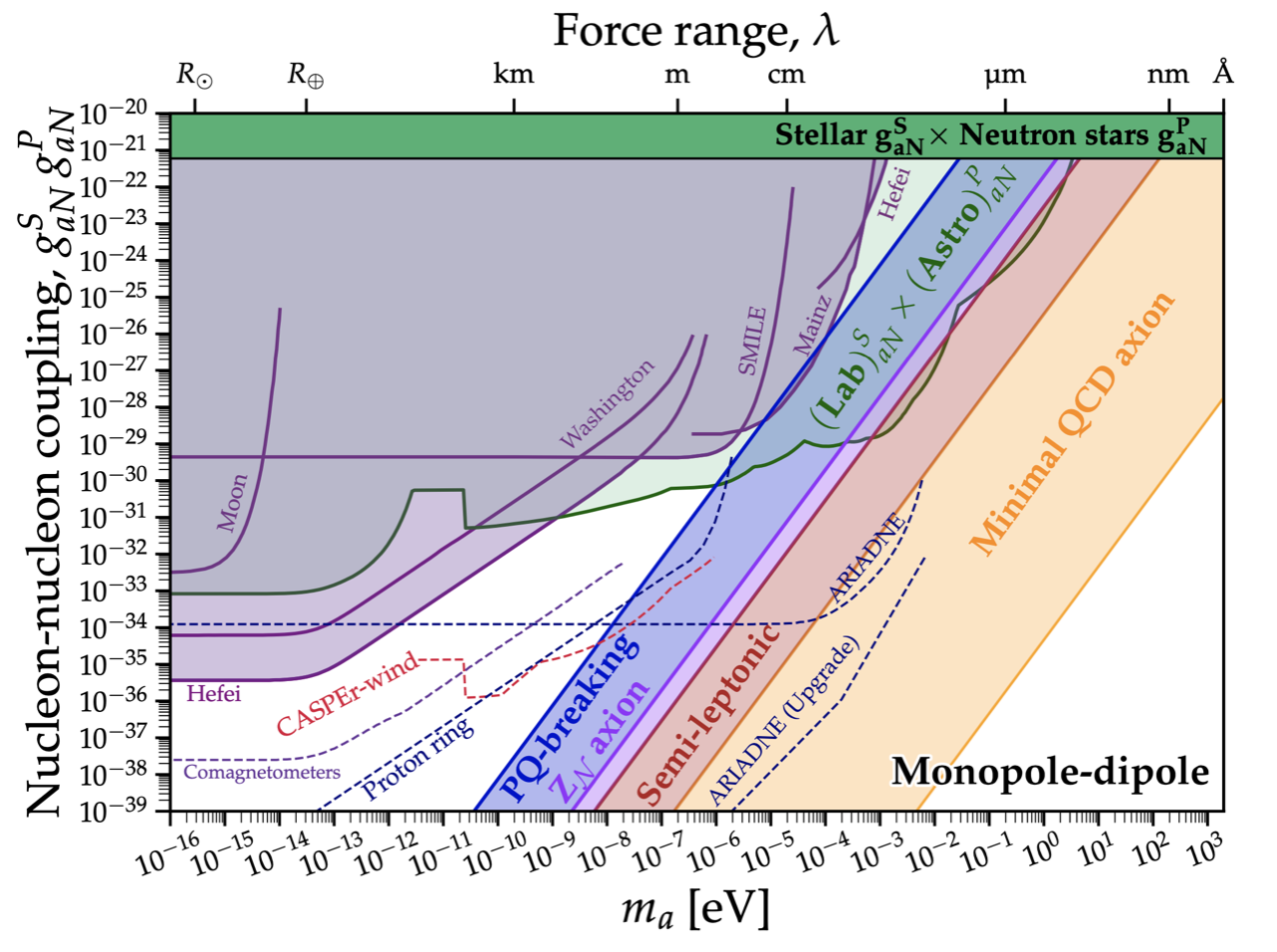}
\caption[]{Parameter space of axion-mediated monopole-dipole forces. 
Figure from Ref.~\cite{DiLuzio:2024ctr}, adapted from \cite{AxionLimits,OHare:2020wah}.}
\label{fig:mondip}       
\end{figure}

\section{Axion-mediated forces}

When both scalar and pseudoscalar couplings are present, three types of non-relativistic axion-mediated potentials arise. Among them, monopole-dipole forces--mediated by scalar-pseudoscalar couplings--are the most promising: monopole-monopole forces are doubly $\theta_{\rm eff}$-suppressed, and dipole-dipole interactions are more strongly spin suppressed. Recently, novel detection concepts based on NMR, such as ARIADNE \cite{Arvanitaki:2014dfa}, have been proposed to probe these effects.  Fig.~\ref{fig:mondip} shows the parameter space in scalar–pseudoscalar coupling vs.~axion mass. 
The yellow band, denoted as ``Minimal QCD axion'',  
is obtained by 
employing 
the value of $\theta_{\rm eff}$ spanning from the CKM contribution 
$\theta^{\rm CKM}_{\rm eff} \sim 10^{-18}$ (lower side)
to the limit imposed by the neutron EDM  
$\theta_{\rm eff} \simeq 10^{-10}$ (upper side). 
However, as mentioned above, the relation between $g^S_{aN}$ and $d_n$ 
is model dependent. Some mechanisms to enhance the scalar axion-nucleon coupling 
(compatibly with EDM bounds) were recently discussed in Ref.~\cite{DiLuzio:2024ctr}. 
These include: 
CPV semi-leptonic operators \cite{Dekens:2022gha} (red),
the $Z_\mathcal{N}$ axion model with modified $m_a$--$f_a$ relation \cite{Hook:2018jle,DiLuzio:2021pxd,DiLuzio:2021gos} (purple), 
and the PQ-breaking electron 
Yukawa scenario \cite{Zhang:2022ykd,Zhang:2023gfu,DiLuzio:2024fyt} (blue).   
These setups extend the conventionally expected range of scalar couplings, even within a QCD axion framework, 
opening up the top-left region of the plot to experimental exploration.

\section{The need for a PQ theory} 

In field-theoretic models, the PQ symmetry is often imposed by hand. A more compelling approach would be to realize it as an accidental symmetry--ideally one protected from high-scale breaking, also addressing the PQ quality problem.
Several proposals attempt this using discrete or continuous gauge symmetries. However, these often rely on UV dynamics that are not testable and introduce symmetries that can appear ad-hoc. A common example is enforcing a $\mathbb{Z}_{10}$  symmetry on a complex scalar $\phi$, hosting the axion as a phase, simply to forbid operators up to $\phi^9$. 
These two drawbacks naturally prompt the following question: can the PQ symmetry emerge from well-motivated gauge structures, such as GUT or flavor symmetries, that are intrinsically connected to the SM?

A possible strategy, relying on the interplay between GUT and flavor gauge symmetries,  
was put forth in Ref.~\cite{DiLuzio:2020qio}, \footnote{A conceptually 
similar approach, based on $\SU(5) \times \SU(3)_{\rm H}$, 
can be found in Refs.~\cite{Berezhiani:1985in,Berezhiani:1990wn}. 
Note, however, that the latter model requires an extra 
symmetry in order to forbid certain PQ-breaking 
cubic operators in the scalar potential. 
The proposal in Ref.~\cite{DiLuzio:2020qio} builds instead upon \cite{Chang:1987hz}, which  
considered a \emph{global} flavor group, thus 
not addressing the PQ 
quality problem and resulting in significantly different phenomenological implications. 
} 
based on the gauge group $\SO(10) \times \SU(3)_f$, where $\SU(3)_f$ denotes the flavor group of SO(10), 
in which the three SM families (plus sterile neutrinos) 
are embedded into a $(16,\3)$ of $\SO(10) \times \SU(3)_f$. 
In this model, the accidental $\U(1)_{\rm PQ}$ 
arises from the interplay of the SO(10) and $\SU(3)_f$ symmetries, 
which furthermore forbid dangerous
PQ-breaking effective operators 
involving large-scale VEVs, 
thus providing a solution to the PQ quality problem.

A remarkable feature inherent to the model is the 
presence of parametrically light fermion fields, 
also known as anomalons,  
introduced to cancel the gauge anomalies of the 
flavor symmetry. 
However, some aspects of the model, 
such as reproducing the SM flavor pattern 
and the phenomenology of the light anomalon fields 
were left unaddressed,  
due to the  complexity of the SO(10) setup. 
In the following, I will report on a recent exploration \cite{DiLuzio:2025jhv} of  a simplified version of the SO(10) model, based on the Pati-Salam gauge group. 
While retaining the key features of the original model, this framework is more tractable from a computational standpoint.

\section{High-quality PQ from the interplay of vertical and horizontal gauge symmetries} 

\begin{table}[t!]
$$\begin{array}{c|c|cc|cc|c|c}
\rowcolor[HTML]{C0C0C0} 
\hbox{Field} & \hbox{Lorentz} &  
G_{\rm PS}
& 
\mathbb{Z}_4 
& \SU(3)_{f_R}  & 
\mathbb{Z}_3 
& {\rm Generations} 
& \U(1)_{\rm PQ} \\ \hline
Q_L & (1/2,0) & (4,2,1) & +i & \1 & +1 & 3 & +3 \\ 
Q_R & (0,1/2) & (4,1,2) & +i & \3 & e^{i 2\pi/3} & 1 & +1 \\ 
\rowcolor{CGray} 
\Psi_{R} & (0,1/2) & (1,1,1) & +1 & \overline{\3} & e^{i 4\pi/3} & 8 & +2 \\
\hline
\Phi & (0,0) & (1,2,2) & +1 & \overline{\3} & e^{i 4\pi/3} & 1 & +2 \\ 
\Sigma & (0,0) & (15,2,2) & +1 & \overline{\3} & e^{i 4\pi/3} & 2 & +2 \\ 
\Delta & (0,0) & (10,1,3) & -1 & \6 & e^{i 4\pi/3} & 1 & +2 \\ 
\chi & (0,0) & (4,1,2) & +i & \overline{\3} & e^{i 4\pi/3} & 1 & -1 \\ 
\xi & (0,0) & (15,1,3) & +1 & \1 & +1 & 1 & \phantom{+}0 \\ 
\end{array}$$
\caption{Field content of the 
Pati-Salam
model and relative transformation properties under 
the Lorentz group, 
$G_{\rm PS}  \times \SU(3)_{f_R}$, its $\mathbb{Z}_4 \times \mathbb{Z}_3$ center,
and the accidental $\U(1)_{\rm PQ}$. 
Exotic fermions, which ensure $\SU(3)_{f_R}$ anomaly cancellation, are highlighted in light gray.}
\label{tab:PSirrep}
\end{table}

In terms of the Pati-Salam group $G_{\rm PS} \equiv \SU(4)_{\rm PS} \times \SU(2)_L \times \SU(2)_R$, 
the minimal model which allows for an accidental $\U(1)_{\rm PQ}$, 
protected also at the EFT level, as well as a consistent flavor breaking pattern, is based on the gauge group
 and $G_{\rm PS} \times \SU(3)_{f_R}$. Its field content is displayed in Table \ref{tab:PSirrep}. 
Although the global (non-abelian) flavor group of $G_{\rm PS}$ is $\SU(3)_{f_R} \times \SU(3)_{f_L}$, 
corresponding to the $\SU(3)$ rotations of the SM fields contained in $Q_L$ and $Q_R$,  we can gauge only the $\SU(3)_{f_R}$ subgroup for phenomenological reasons. 
Indeed, if the $\SU(3)_{f_L}$ factor were also gauged, the scales of flavor and electroweak (EW)
breaking would be related, leading either to 
high-scale breaking of EW symmetry or 
to unwanted EW-scale flavor gauge bosons. 

On top of the SM fermions (plus right-handed neutrinos), the model also features a set of anomalon fields, 
$\Psi_R$, which are introduced in order to cancel the $\SU(3)_{f_R}^3$ gauge anomaly.
A remarkable feature of the anomalon sector, in common with the $\SO(10) \times \SU(3)_{f}$ model \cite{DiLuzio:2020qio}, 
is that the $\Psi_R$ fermions are massless at the renormalizable level, 
which is closely connected to the existence of the accidental $\U(1)_{\rm PQ}$. 
Non-renormalizable operators will lift the mass of the anomalons, which represent a low-energy signature of this 
setup. 

Coming to the scalar sector of the model, the fields $\Phi$, $\Sigma$ and $\Delta$ are required 
to reproduce 
SM fermion masses and mixings in a renormalizable Yukawa sector.
The field $\chi$ is introduced 
for rank reasons, 
in order to fully break both the $\mathrm{U}(1)_{B-L}$ of $\SU(4)_{\rm PS}$ and the accidental $\U(1)_{\rm PQ}$ symmetry at high energies, 
well above the EW scale. Finally, the real scalar $\xi$ is an auxiliary field that is required in the scalar potential in order to generate the accidental $\U(1)_{\rm PQ}$.

\begin{figure}[t!]
\centering
\includegraphics[width=16.cm]{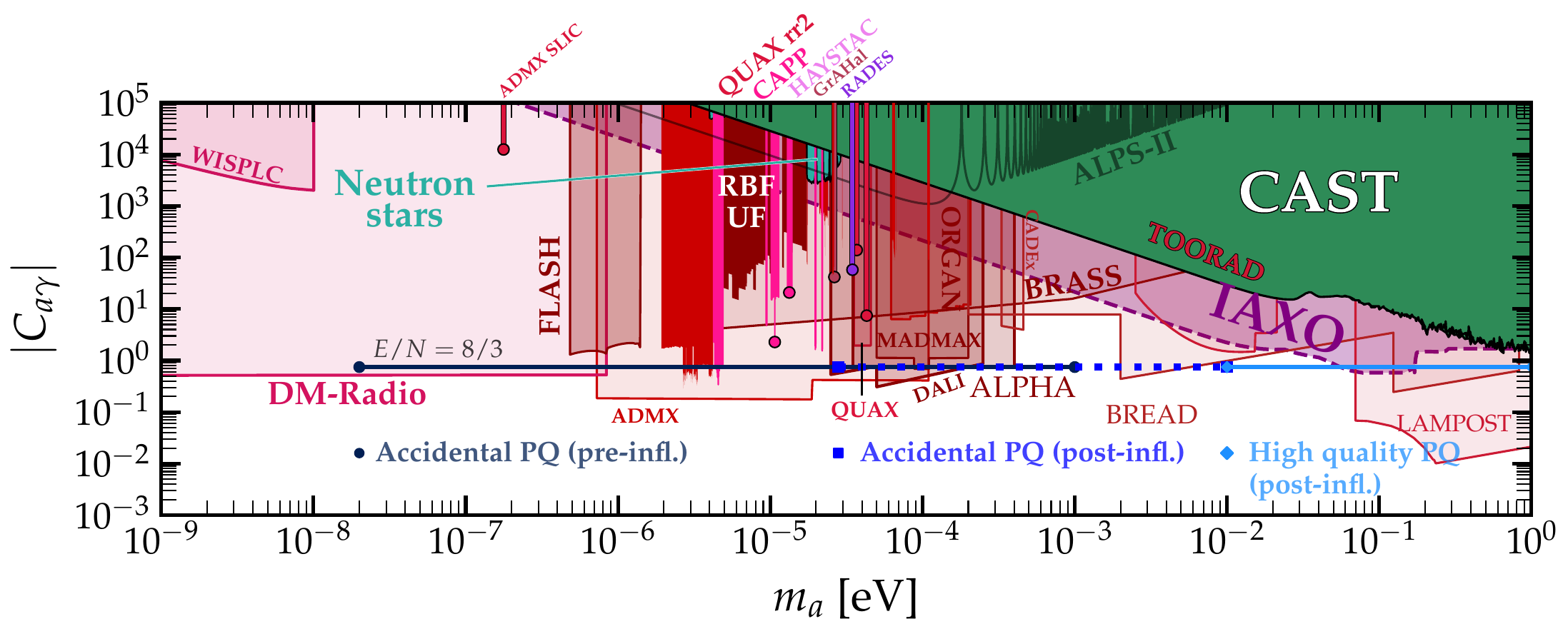}
\caption[]{Axion-photon coupling 
as a 
function of the axion mass, in the model of Ref.  \cite{DiLuzio:2025jhv}. 
Current limits and future sensitivities adapted from~\cite{AxionLimits}.
A high-quality PQ symmetry is obtained for $m_a \gtrsim 0.01\,\mathrm{eV}$ (light blue line). 
}
\label{fig:axionphoton}       
\end{figure}

While we refer the reader to \cite{DiLuzio:2025jhv} for the details, we summarize here the main results: 
\begin{itemize}
\item We analyzed the model's ability 
to address the PQ quality problem and we showed that it can non-trivially reproduce the SM flavor structure.
\item The main predictions of the model, regarding axion phenomenology, 
are summarized in \fig{fig:axionphoton}, where three 
distinct 
axion mass 
windows are identified: 
$i)$ $m_a \in [2 \times 10^{-8}, 10^{-3}]\,\mathrm{eV}$, 
$ii)$ $m_a \in [2.8 \times 10^{-5}, 0.01]\,\mathrm{eV}$ 
(corresponding to an accidental $\U(1)_{\rm PQ}$, 
respectively in the 
pre-inflationary and post-inflationary PQ breaking scenario), and $iii)$ $m_a \gtrsim 0.01\,\mathrm{eV}$
(associated with a high-quality $\U(1)_{\rm PQ}$ in the post-inflationary PQ breaking scenario).

\item 
We computed the anomalon–neutrino mixing (relevant below the EW scale), and showed that high-quality PQ symmetry predicts sub-eV anomalons. 
As a result, they contribute to the dark radiation of the universe. While for intermediate-scale 
values of $f_a$, the anomalons can be heavier (keV scale) 
and thus contribute to dark matter. 

\item We studied the cosmological production of anomalons in the early universe, considering their production via flavor interactions, neutrino mixing and effective Yukawa-like operators. Current constraints on the effective number of relativistic species, $\Delta N_{\rm eff}$, from Planck'18 
data were applied to constrain the parameter space. We further highlighted how future precision measurements of 
$\Delta N_{\rm eff}$ could serve as a low-energy probe of the UV dynamics underlying the solution to the PQ quality problem. 

\end{itemize}

\section{Conclusions} 

To conclude, imperfect axions lead to interesting phenomenology. CPV and PQ breaking induce scalar axion couplings, enhancing axion-mediated forces that can be tested in the current generation of axion experiments. In this context, the axion provides a low-energy portal to new sources of CPV beyond CKM. I also presented a new solution to the PQ quality problem based on the interplay of GUT and flavor symmetries. It predicts a relatively heavy axion 
$m_a \gtrsim 0.1$ eV and parametrically light anomalons, as potential low-energy signatures of the mechanism enforcing PQ quality.

\section*{Acknowledgments}

Work 
supported
by the European Union -- Next Generation EU and
by the Italian Ministry of University and Research (MUR) 
via the PRIN 2022 project n.~2022K4B58X -- AxionOrigins.

\section*{References}

\bibliographystyle{unsrtnt}
\bibliography{biblio}

\end{document}